# Design of Voltage Pulse Control Module for Free Space Measurement-Device-Independent Quantum Key Distribution

Sijie Zhang, Nan Zhou, Fanshui Deng and Hao Liang

*Abstract*—Measurement-Device-Independent Quantum Key Distribution (MDIQKD) protocol has been proved that it is unaffected by all hacking attacks, and ensures the security of information theory even when the performance of single-photon detectors is not ideal. Fiber channel has been used by the previous MDIQKD experimental device. However, the signal attenuation increases exponentially as the transmission distance increases. In order to overcome this, we regard free space as the channel of signal transmission, and the signal attenuation increases square as the transmission distance increases (regardless of the atmospheric scattering), which can effectively reduce the signal attenuation trend. In order to implement the free space MDIQKD experiments, a modulation module is needed to modulate the wide pulse chopping, decoy-state, normalization, phase encoding and time encoding. In this paper, we present the design of the Voltage Pulse Control Module for the free space MDIQKD.

*Index Terms*—Voltage pulse control, electronics design, free space MDIQKD.

## I. Introduction

COMPARED with classical communication, the greatest advantage of quantum communication is security. In the information society where technology is in the same interest and prosperity, people are increasingly demanding security, especially in areas and sectors that require high security, such as military and financial services and so on. Quantum communication is almost the only choice for future secure communications. There are many research directions for quantum communication, and quantum cryptography is the most important of which is the first practical field.

Measurement-device-independent quantum key distribution (MDIQKD) protocol [1], [2], which is immune to all hacking attacks on detection, guarantees the security of information theoretically even with single-photon detectors, whose performance is not perfect. Fiber channel is used by the previous MDIQKD experimental device. However, the signal attenuation increases exponentially along with the transmission distance increases. But using free space as the channel for signal transmission, with the signal attenuation increases squarely (without considering the scattering of the atmosphere), the signal attenuation trend can be effectively reduced.

In order to implement free space MDIQKD experiments, a modulation module is needed to modulate the wide pulse chopping, decoy-state, normalization, phase encoding and time encoding.

In this paper, we report our work on the design of voltage pulse control module, focusing on the realization of pulse control using a wideband amplifier offering high dynamic range, on the prevention of waveform distortion, and insufficient drive capacity. The voltage pulse control module designed for the modulation of the optical module of the free space MDIQKD has fulfilled the requirements of the laboratory environment and is able to now be applied in an actual experiments. And we did conduct joint debugging of the entire physical experiment platform.

Manuscript received June 1, 2018. This work was supported in part by the National Fundamental Research Program under Grants 2011CB921300, 2013CB336800, and 2011CBA00300, the National Natural Science Foundation of China, the Chinese Academy of Science, the Quantum Communication Technology Co., Ltd., Anhui, and the Shandong Institute of Quantum Science and Technology Co., Ltd.

S. J. Zhang, F. S. Deng, and H. Liang are with the State Key Laboratory of Particle Detection and Electronics, University of Science and Technology of China, Hefei, Anhui, 230026, China (e-mail: zsj2015@mail.ustc.edu.cn; dfsh@mail.ustc.edu.cn; simonlh@ustc.edu.cn).

N. Zhou, was with the State Key Laboratory of Particle Detection and Electronics, University of Science and Technology of China, Hefei 230026, China. He is now with Meiya Optoelectronics Technology Co., Ltd, Hefei, Anhui, 230088, China (e-mail: zhoun@mail.ustc.edu.cn;).

## II. DESIGN

### A. System Structure

The voltage pulse control module is used in the quantum cryptography system, with the microcontroller and DAC chip as the core. The modulation signal is generated under the control of the upper computer for modulating the light pulse that reaches the modulator.

The main function of the voltage pulse control module [3] is as a modulation circuit, with a total of 5 Digital to Analog Converter (DAC) outputs. Respectively, the wide pulse chopper DAC has two adjustable amplitudes outputs(AC1, AC2), the decoy-state control DAC has four adjustable amplitudes outputs (AD1, AD2, AD3, AD4) and the normalized control DAC has two adjustable amplitudes (AU1, AU2). There are two adjustable amplitudes outputs (AP1, AP2) for the phase encoding control DAC and two adjustable amplitudes outputs (AT1, AT2) for the time encoding control DAC. Figure. 1 illustrate the structure of the electronics system briefly.

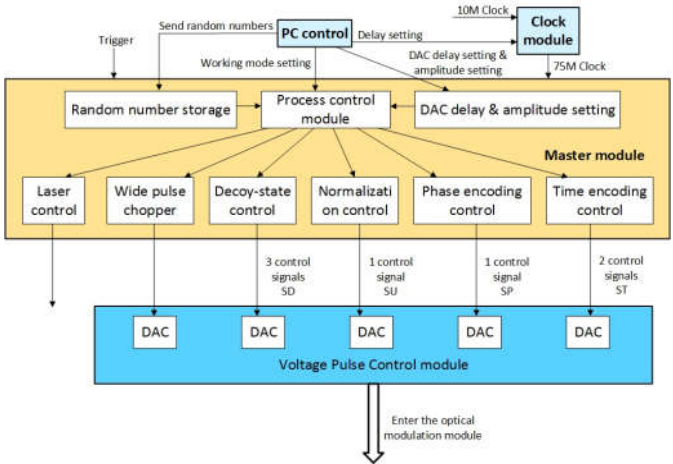

Fig. 1. The simplified structure of the electronics system

Specifications require that the output voltage of all 5 DACs can be adjustable from 0V to 6V in 0.05V steps. And all 5 DACs have adjustable delay times of ± 15ns in 100ps steps. The rising edge of the DAC output is about 1ns, and the maximum output amplitude can reach 7V with a 50Ω impedance. Therefore, the power requirements of voltage pulse control module are relatively high. However, with the increase of power, we can get a flatter pulse, which can better serve the subsequent optical modulation module. Later, we will be going to implement the system with a clock frequency of GHz. We expect the security key rate can be further increased by increasing the clock frequency.

### B. Hardware

In the voltage pulse control module design, the output module of previous design for the fiber channel MDIQKD uses THS3201 [4], which is a wide gain bandwidth, high speed current feedback amplifier of Texas Instruments (TI), after two stages of amplification, then outputs the voltage. But, there is still a need for a RF power amplifier module independent of the voltage pulse control module before it can be applied to the MDIQKD system. However, in the new design, we put the RF power amplifier directly on the circuit board, so that we can get enough power and stable voltage output only by using the one stage operational amplifier and the RF power amplifier. This eliminates the need for an independent RF power module and at the same time we get the output voltage what we need. The block diagram of the new design voltage pulse control module is shown in the Figure 2.

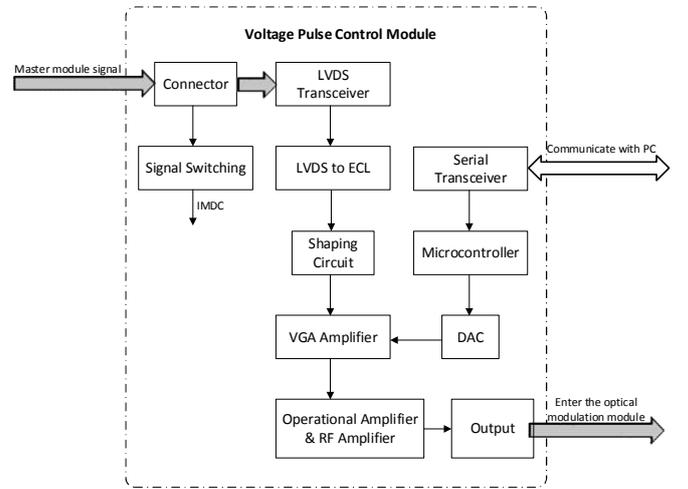

Fig. 2. Block diagram of the voltage pulse control module

The signal flow of the entire module is described as follows. As is seen in the Figure 2, the Low Voltage Differential Signal (LVDS) enters the voltage pulse control module through the connector and then is transmitted to the LVDS transceiver after signal selection. After level shifting, the signal level is converted to Emitter Coupled Logic (ECL), and then the signal flows into the shaping circuit and Voltage-controlled Gain Amplifier (VGA) amplification circuit. The control system sends a command to the voltage pulse control module via the serial port to modulate the DAC outputs, which in turn control the VGA outputs, thereby modulating the output voltage pulse.

## III. TEST RESULTS

### A. Laboratory environment test

The tests are in progress and we are also starting the part work on the design of voltage pulse control module for the GHz clock frequency. The test method we used currently is that the input signal enters the voltage pulse control module from the master module of fiber channel system via the connector. Then the control system sends a command to the voltage pulse control module through the serial port, and the output signal is connected to the oscilloscope, which is the LeCroy WaveRunner 640Zi oscilloscope [5] with a 4GHz bandwidth and a sampling rate of 40GS/s, for observing the output voltage waveform.

With a 50Ω load, the output waveform amplitude is up to 10Vpp. The output frequency meets the experimental requirements, and the experimental results show that the design index is satisfied.

### B. Physics experiment platform test

The results of the current preliminary tests are shown in Figure 3. More detailed results of the joint adjustment with the physics experiment platform will be provided recently.

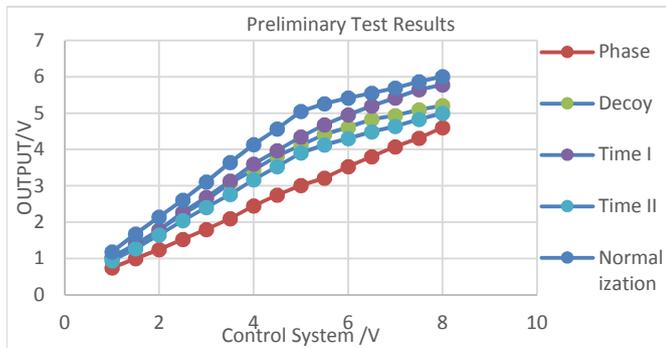

Fig. 3. Preliminary test results

The overall design and production of the prototype have been completed in April 2018 and are currently undergoing joint deployment with the physical platform. The entire electronics system will be completed in July 2018. After that, we will devote all our energy to the design of the voltage pulse control module with a clock frequency of GHz based on this. By increasing the clock frequency, we believe that we can further increase the security key rate.

### IV. CONCLUSION

This paper describes the design of one electronic part for quantum cryptographic system - voltage pulse control module.

For the voltage pulse generator, although it has initially reached the working requirements and has been applied to the actual experiment through the system joint debugging, there are still some areas for improvement. The increase in voltage configuration speed and elimination of signal edge jitter is the requirement for the next step.